\documentclass[]{article}

\usepackage{amsmath}
\usepackage{amsfonts}
\usepackage{amssymb}
\usepackage{amsthm}
\usepackage{a4}
\usepackage{amssymb,amsmath}
\usepackage{graphicx}
\usepackage{bbm}
\usepackage{bm}

\newcommand{\vP}{\bm{F}_{\rm P}}
\newcommand{\vlat}{\bm{F}_{\rm{lat}}}
\newcommand{\vLAT}{\bm{F}_{\rm{LAT}}}
\newcommand{\vLoad}{\bm{F}_{\rm{load}}}
\newcommand{\vLOAD}{\bm{F}_{\rm{LOAD}}}
\newcommand{\vF}{\bm{F}}
\newcommand{\vef}{\hat{\mathbf{f}}}
\newcommand{\vei}{\hat{\mathbf{i}}}
\newcommand{\vS}{\bm{F}_{\rm S}}
\newcommand{\vC}{\bm{F}_{\rm C}}
\newcommand{\vN}{\bm {F}_{\rm N}}
\newcommand{\vW}{\bm{W}}
\newcommand{\vex}{\hat{\bm{x}}}
\newcommand{\vn}{\hat{\bm{n}}}

\begin{document}

\title{\sf Physics of Skiing:\\
The Ideal--Carving Equation and Its Applications}

\author{U. D. Jentschura and F. Fahrbach\\[1ex]
{\em Universit\"at Freiburg, Physikalisches Institut,}\\
{\em Hermann--Herder--Stra\ss{}e 3,}\\
{\em 79104 Freiburg im Breisgau, Germany}}

\maketitle

\normalsize
\begin{abstract}
Ideal carving occurs when a snowboarder or 
skier, equipped with a snowboard or carving skis,
describes a perfect carved turn in which the edges of the ski
alone, not the ski surface, describe the trajectory followed
by the skier, without any slipping or skidding.
In this article, we derive the ``ideal-carving'' equation
which describes the physics of a carved turn under ideal
conditions. The laws of Newtonian classical mechanics are applied.
The parameters of the ideal-carving equation
are the inclination of the ski slope, the acceleration of 
gravity, and the sidecut radius of the ski. The variables
of the ideal-carving equation
are the velocity of the skier, the angle between the 
trajectory of the skier and the horizontal, and the 
instantaneous curvature radius of the skier's trajectory.
Relations between the slope inclination and the velocity 
range suited for nearly ideal carving are discussed,
as well as implications for the design of carving skis
and snowboards. 
\end{abstract}

%
%
\section{Introduction} 

The physics of skiing has recently been described in a 
rather comprehensive book~\cite{LiSa1996}
which also contains further references of interest. The current article is 
devoted to a discussion of the forces acting on a skier or 
snowboarder, and to the derivation of an equation which describes
``ideal carving'', including applications of this concept in 
practice and possible speculative implications for the design of technically
advanced skis and snowboards. 
In a carved turn, it is the bent, curved edge of the ski or snowboard
which forms some sort of ``railroad track'' along which the 
trajectory of the curve is being followed, as opposed to 
more traditional curves which are triggered by deliberate
slippage of the bottom surface of the ski relative to the snow.

\begin{figure}[htb!]
\begin{center}
\begin{minipage}{13.0cm}
\begin{center}
\vspace*{1cm}
\includegraphics[scale=0.68]{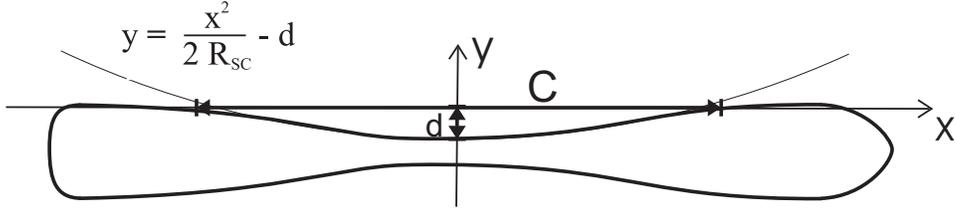}
\caption{\label{fig1}
Overhead view on a typical carving ski. The tip of the ski 
points to the right.
The geometry of the carving equipment involves the contact length $C$,
the sidecut $d$ as well as the sidecut radius 
$R_{\rm SC}$. For a more traditional ski, the sidecut
would be negligibly small, resulting in 
a very large sidecut radius.}
\end{center}
\end{minipage}
\end{center}
\end{figure}

The edges of traditional skis have a nearly straight-line 
geometry. By contrast, carving skis (see figure~\ref{fig1}) 
have a manifestly nonvanishing 
sidecut\footnote{Materials used in the construction of carving skis
have to fulfill rather high demands because the strongest
forces act on the narrowest portion of the ski. Yet at the
same time, undesired vibrations of the shovel and tail
of the ski have to be avoided, so that the materials have to be
rigid enough to press both shovel and tail firmly onto the
snow and absorb as well the strong lateral forces exerted during
turns.}. Typical carving skis have a sidecut radius of the order
of $16\,{\rm m}$ at a chord length of $170\,{\rm cm}$.
However, parameters used in various models
may be adapted to the intended application:
e.g., models designed for less narrow curves may have a sidecut radius
$R_{\rm SC}$ of about $19.5\,{\rm m}$ at a length of $180\,{\rm cm}$.
Skis suited for very narrow slalom curves
have an $R_{\rm SC}$ of
about $14.5\,{\rm m}$ at a length of up to $164\,{\rm cm}$.
A typical version suited for off-piste freestyle skiing
features a typical length of $185\,{\rm cm}$
at a sidecut radius of $25.3\,{\rm m}$.

%
%

The sidecut has been introduced first into the world 
of snow-related leisure activity by the snowboard
where the wider profile of the instrument allowed for 
a realization of a rather marked sidecut without 
stringent demands on the materials used.
The carved turn is the preferred turning procedure on 
a snowboard. Skidding should be avoided in competitions
as far as possible because it necessarily 
entails frictional losses. In practical situations,
carving skiers and snowboarders may realize 
astonishingly high tilt angles $\phi$ 
in the range of $[60^o, 80^o]$ (the ``tilt angle'' 
is the angle of line joining the board 
and the center-of-mass of the skier or boarder
with the normal to the surface of the ski slope).
In snowboarding, the transition from a right to a left
turn is then often executed by jumping, with the idea
of ``glueing together'' the trajectories of two perfect 
carved turns, again avoiding frictional losses.
The sidecut radius $R_{\rm SC}$ of a snowboard may assume values as low as
6--8$\,{\rm m}$ at a typical chord length between
$150\,{\rm cm}$ and $160\,{\rm cm}$.

This paper is organized as follows: In section~\ref{sToward},
we proceed towards the derivation of the ``ideal-carving equation'',
which involves rather elementary considerations that we chose to 
present in some detail. 
The geometry of sidecut radius is discussed in section~\ref{ssSidecut},
and we then proceed from the simplest case of 
the forces acting along
a trajectory perpendicular to the fall line (section~\ref{ssPerp})
to more general situations~(section~\ref{ssGeneral}).
The projection of the forces in directions parallel and 
perpendicular to the skier's trajectory lead to the 
concept of ``effective weight''~(section~\ref{ssWeight}).
The inclusion of a centrifugal force acting in a curve
allows us to generalize this concept (section~\ref{ssActing}).
In section~\ref{sICE}, we discuss the ``ideal-carving equation'',
including applications. 
The actual derivation of the ``ideal-carving equation'' in 
section~\ref{ssDerivation} is an easy exercise, 
in view of the preparations made in section~\ref{sToward}.
An alternative form of this equation has already appeared 
in~\cite{LiSa1996}; in the current article we try to reformulate
the equation in a form which we believe is more suitable for 
practical applications. The consequences which follow from
the ``ideal-carving equation'' 
are discussed in 
section~\ref{ssGraphical}. Finally, we mention some
implications  for the design of skis and boards 
in section~\ref{sConclusions}.

%
%
\section{Toward the Ideal--Carving Equation}
\label{sToward}

%
%
\subsection{The Sidecut--Radius}
\label{ssSidecut}

According to figure~\ref{fig1}, 
an approximate formula for the
smooth curve described by the edge of the ski is 
$y \approx x^2/(2 R_{\rm SC}) - d$. This relation implies
$d \approx C^2/(8 R_{\rm SC})$.
Therefore, we approximately have
\begin{equation} 
\frac{{\rm d}^2y}{{\rm d}x^2} \,  \approx \, \frac{1}{R_{\rm SC}}\,,
\end{equation}
where $R_{\rm SC}$ is the sidecut radius of the ski.
We then have
\begin{equation}
y  \, \approx \,  \frac{x^2}{2 \, R_{\rm SC}} - d
\end{equation} 
along the edge of the ski.
For $x = C/2$ we have $y \approx 0$ and therefore
\begin{equation}
\label{RSC}
d \approx \frac{C^2}{8 \, R_{\rm SC}}\,,
\qquad 
R_{\rm SC} \approx \frac{C^2}{8 \, d}\,.
\end{equation} 
A typical carving ski with a chord length 
of $160\,{\rm cm}$ may have a sidecut radius
as low as $R_{\rm SC} = 14\,{\rm m}$. Estimating the contact length 
to be about $80\,\%$ of the chord length,
we arrive at $d \approx~1.8\,{\rm cm}$.  
The maximum width of a ski (at the 
end of the contact curve) is of the order of 
$10\,{\rm cm}$, so that a sidecut of about $1.8\,{\rm cm}$
will lead to a minimum width which is roughly 
two thirds of the maximum width, leading to high structural demands
on the material. 

A typical snowboard has a length of about $168\,{\rm cm}$ and 
a width of about $25\,{\rm cm}$. 
The contact length is $125\,{\rm cm}$ in a typical case.
A sidecut of
\begin{equation}
d \, \approx \, \frac{C^2}{8 \, R_{\rm SC}} \approx\, 2.4\,{\rm cm}
\end{equation} 
results, which means that for a typical snowboard,
the relative difference of the minimum width to 
the maximum width is only $20\,\%$. 

As the ski describes a carved turn, 
the radius of curvature $R$ along the trajectory varies with the 
angle of inclination $\phi$ of the normal to the 
ski slope with the normal to the ski surface.
For $\phi \approx 0$, we of course have $R \approx R_{\rm SC}$.
An elementary geometrical consideration shows that the 
effective sidecut $d'$ at inclination $\phi$ is given by
\begin{equation}
d' = \frac{d}{\cos\phi}\,.
\end{equation}
The inclination-dependent sidecut-radius is therefore
\begin{equation} 
\label{Rphi}
R(\phi) = \frac{C^2}{8d'} = R_{\rm SC} \, \cos\phi \leq R_{\rm SC}\,.
\end{equation}

%
%
\subsection{Trajectory Perpendicular to the Fall Line}
\label{ssPerp}

The angle of inclination of the ski slope against the horizontal 
is $\alpha \in [0^o, 90^o]$. We assume the skier's trajectory to be 
exactly horizontal. The work done by the gravitational force
vanishes, and the skier, under ideal conditions, neither 
decelerates nor accelerates.

%
%
\begin{figure}[htb!]
\begin{center}
\begin{minipage}{14.0cm}
\begin{center}
\vspace*{1cm}
\includegraphics[scale=0.6]{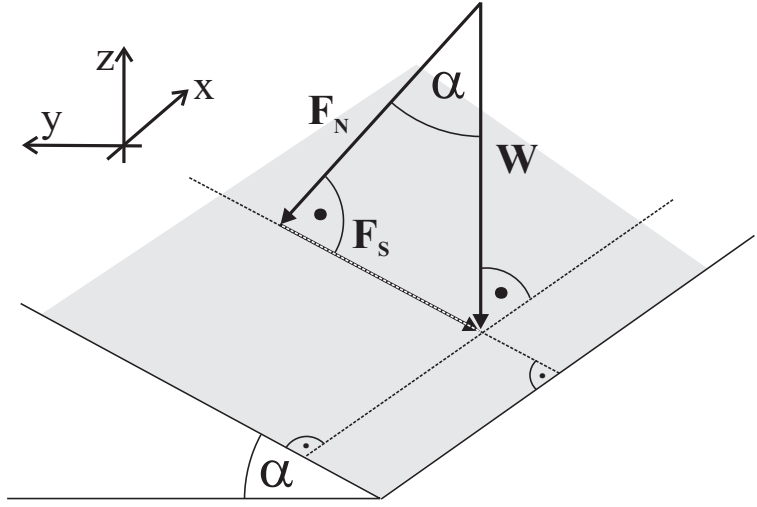}
\caption{\label{fig2}
The skiing trajectory is indicated as parallel to a horizontal 
that lies in the plane described by the ski slope. That plane, in turn,
is inclined against the horizontal by an angle $\alpha$.
Of course, the angle $\alpha$ is equal to the angle of $\bm{F}_{\rm N}$ with
$\bm{W}$.
The force exerted on the skier in the gravitational
field of the Earth is $\vW$. It may be decomposed into 
a component $\vN$ perpendicular to the plane described 
by the ski slope, and a component $\vS$ lying in the 
plane of the slope.}
\end{center}
\end{minipage}
\end{center}
\end{figure}

With axes as outlined in figure~\ref{fig2}, we have
\begin{equation}
\vW = 
\left( \begin{array} {c} 0 \\ 0 \\ - \, m \, g \end{array} \right)\,,
\qquad \|\vW\| = m\,g\,.
\end{equation}
In order to calculate $\vN$, we project $\vW$ onto the 
unit normal of the ski slope,
\begin{equation}
\label{defFN}
\vn  =  -\frac{\vN}{\|\vN\|}  =  
\left(\begin{array}{c} 0 \\ -\sin\alpha \\ \cos\alpha \end{array}\right)\,,
\qquad
\vN  =  (\vn \cdot \vW) \, \vn\,.
\end{equation}
This leads to the following representation,
\begin{equation}
\vN = m \, g \,	\left(\begin{array}{c} 
0 \\ \sin\alpha\,\cos\alpha \\ -\cos^2\alpha \end{array}\right)\,, 
\qquad
\| \vN \| = m\,g\,\cos\alpha\,.
\end{equation}
$\vS$ has the representation:
\begin{equation}
\label{defFS}
\vS =  \vW \, - \, \vN = m\,g \,
\left(\begin{array}{c} 0 \\ 
-\sin\alpha\cos\alpha \\ 
-\sin^2\alpha \end{array}\right)\,, \qquad
|\vS| = m\,g\,\sin\alpha\,.
\end{equation}

%
%
\subsection{More General Case} 
\label{ssGeneral}

\begin{figure}[htb!]
\begin{center}
\begin{minipage}{14.0cm}
\begin{center}
\vspace*{1cm}
\includegraphics[scale=0.6]{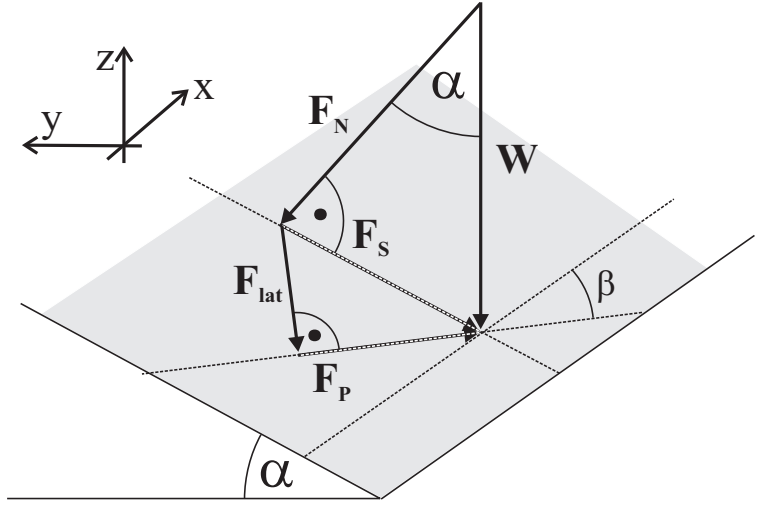}
\caption{\label{fig3}
This figure illustrates vector decomposition in a more general case.
The skier's trajectory describes an angle 
$\beta$ with the horizontal and is explicitly indicated in 
the figure by the straight line parallel to $\vP$. The force
$\vS$, which is acting inside the slope plane, is
decomposed into a lateral force $\vlat$ and 
a force $\vP$ which is parallel to the skier's trajectory and 
leads to an acceleration.
Of course, the force $\vN$ is perpendicular to the slope.}
\end{center}
\end{minipage}
\end{center}
\end{figure}

In a more general case, the angle of the skier's 
trajectory with the horizontal is denoted by $\beta$.
Within a right curve, $\beta$ varies from $0^o$ to 
$180^o$, whereas within a left-hand curve, 
$\beta$ varies from an initial value of 
$180^o$ to a final value of $0^o$. 
The elementary geometrical considerations follow 
from figure~\ref{fig3}.

%
%

For a skier's trajectory directly in the 
fall line, we have $\beta = 90^o$, $\vlat=0$ and 
$\vS=\vP$, resulting in maximum acceleration along the 
fall line.
Let $\vef$ be a unit vector tangent to the 
skier's trajectory, i.e.~ $\vef \, || \, \vP$,
and $\| \vef \| = 1$.
The direction of $\vP$ is displayed in figure~\ref{fig3}.
An analytic expression
for $\vef$ can easily be obtained by starting 
from a unit vector $\vex$ as implied by the conventions 
used in figure~\ref{fig3}.
We first rotate $\vex$ about the $z$-axis by an angle $-\beta$,
and obtain a vector $\vei$. A further rotation about 
the $x$-axis by an angle $+\alpha$ leads
to the vector $\vef$.

Of course, a rotation of $\vex$ by $-\beta$ leads to
$\vei = (\cos\beta, - \sin\beta, 0)$.
The rotation of $\vei$ about the $x$-axis by an angle $\alpha$,
which ``rotates $\vei$ into the ski slope'', is expressed
as a rotation matrix,
\begin{equation}
R = \left(
\begin{array}{ccc} 
1 & 0& 0 \\ 
0 & \cos\alpha & -\sin\alpha \\
0 & \sin\alpha & \sin\alpha
\end{array}\right) \quad 
\Rightarrow \quad
\vef = R \cdot \vei = 
\left(\begin{array}{c} \cos\beta \\ 
- \cos\alpha\sin\beta \\ 
-\sin\alpha\sin\beta \end{array}\right).
\end{equation}
Then, by projection,
\begin{equation}
\vP = (\vef \cdot \vS) \,\vef = - m\,g \,
\left(\begin{array}{c} 
-\sin\alpha\,\sin\beta\,\cos\beta  \\
\sin\alpha\,\cos\alpha\,\sin^2\beta  \\
\sin^2\alpha\,\sin^2\beta
\end{array}\right)\,,
\end{equation}	
and the modulus of the vector $\vP$ is 
\begin{equation}
\|\vP\| = \|\vS\| \,|\sin\beta| \, = \, m\,g\, \sin\alpha \, |\sin\beta|\,.
\end{equation}
The force $\vlat$ perpendicular to the track direction is 
then easily calculated with the help of equation~(\ref{defFS}),
\begin{equation}
\label{defFlat}
\vlat = \vS - \vP = - m\,g \,
\left(\begin{array}{c} 
\sin\alpha\,\sin\beta\,\cos\beta  \\
\sin\alpha\,\cos\alpha\,\cos^2\beta  \\
\sin^2\alpha\,\cos^2\beta
\end{array}\right),\\
\end{equation}
resulting in
\begin{equation}
\| \vlat \| = \|\vS\| \, |\cos\beta| = m\,g\, \sin\alpha \, |\cos\beta|\,.
\end{equation}
This force has to be compensated by the snow, under the approximation
of vanishing slippage.

%
%
\subsection{Effective Weight}
\label{ssWeight}

According to figure~\ref{fig3},
the gravitational force may be decomposed as
$\vW = \vN + \vlat + \vP$. The force $\vP$ simply accelerates the 
skier. The force $\vLoad = \vN + \vlat$ therefore has to be compensated by
the snow. Therefore, to avoid slippage, the skier should 
balance her/his weight in such a way that his/her center-of-mass
is joined with the ski along a straight line parallel to 
the direction given by the effective weight vector $\vLoad$,
\begin{equation}
\label{defload}
\vLoad = \vW - \vP = \vN + \vlat\,.
\end{equation}
We immediately obtain
\begin{equation}
\vLoad = m\,g \, \left(\begin{array}{c} 
-\sin\alpha\,\sin\beta\,\cos\beta  \\
\sin\alpha\,\cos\alpha\,\sin^2\beta  \\
\sin^2\alpha\sin^2\beta - 1
\end{array}\right)\,, \quad
\| \vLoad \| = 
\sqrt{\vN^2 + \vlat^2} = m\,g\, \sqrt{ \sin^2\alpha\sin^2\beta + 1 }\,.
\end{equation}

%
%
\begin{figure}[htb!]
\begin{center}
\begin{minipage}{14.0cm}
\begin{center}
\vspace*{1cm}
\includegraphics[scale=0.6]{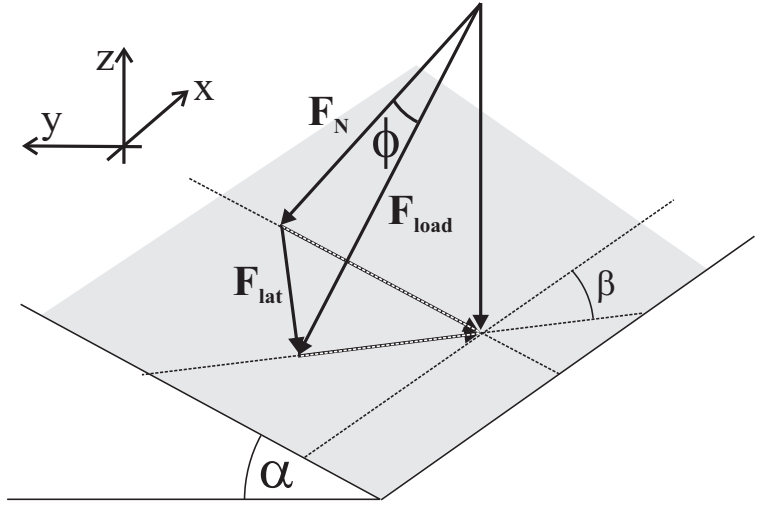}
\caption{\label{fig4}
The skier's trajectory is as indicated in figure~\ref{fig3}.
The effective weight of the skier is $\vLoad$. 
We remember that $\vN$ is perpendicular 
to the ski slope. In order to avoid slippage, the line
joining the ski boots and the 
skier's center-of-mass must be parallel
to $\vLoad$. 
The tilt angle of the skier with the normal to the ski slope is equal to
the angle of $\bm{F}_{\rm{load}}$ with $\bm {F}_{\rm N}$; this angle is
denoted by the symbol~$\phi$ in the text.}
\end{center}
\end{minipage}
\end{center}
\end{figure}

We call $\phi$ the angle of inclination or tilt angle.
For obvious reasons, we assume this angle to be in the range 
$\phi \in [0^o, 90^o]$.
The angle $\phi$ also enters in the inclination-dependent
sidecut radius as given in equation (\ref{Rphi}).
We now investigate the question
how $\phi$ is related to $\alpha$ and $\beta$. 
To this end, we first calculate
\begin{equation}
\label{cosphi}
\cos\phi = \frac{\vLoad \cdot \vN }{\| \vLoad \| \, \| \vN \|} 
= \frac{\cos\alpha}{\sqrt{\cos^2\alpha + \cos^2\beta \, \sin^2\alpha}}\,.
\end{equation}
We now apply the identity $\tan[\cos^{-1}(x)] = \sqrt{1-x^2}/x$
to both sides of this equation. The left-hand side becomes
just $\tan \phi$, and the right-hand side simplifies
to $\tan\alpha\, |\cos\beta|$.

Alternatively, we observe that $\vN$, $\vlat$ and
$\vLoad$ are vectors in one and the same plane. 
Because $\vN$ is perpendicular to the ski slope
and $\vlat$ is a vector in the plane formed by the slope,
both vectors are at right angles to each other (see figure~\ref{fig4}).
We immediately obtain
\begin{equation} 
\label{defphi}
\tan\phi = \frac{\| \vlat \| }{\| \vN \|} = 
\frac{m\,g\, \sin\alpha \, |\cos\beta|}{m\,g\,\cos\alpha}
= \tan\alpha \, |\cos\beta|\,.
\end{equation}	

%
%
\subsection{Forces Acting in a Curve}
\label{ssActing}

\begin{figure}[htb!]
\begin{center}
\begin{minipage}{14.0cm}
\begin{center}
\vspace*{1cm}
\includegraphics[scale=0.6]{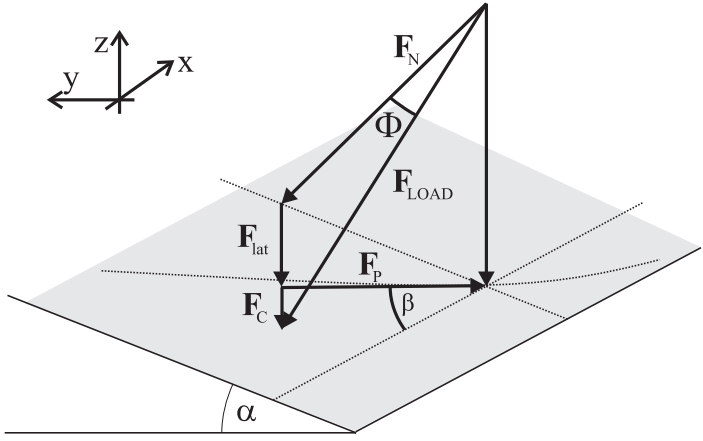}
\caption{\label{fig5}
The total lateral force $\vLAT$ in a curve is the 
sum of $\vlat$ given in (\ref{defFlat}) and 
the centrifugal force $\vC$ defined in (\ref{defFC}).
This leads to a modified effective weight $\vLOAD$.
For the centrifugal force to act as shown in the figure,
the skier's trajectory is required to describe a left-hand
curve, as displayed.}
\end{center}
\end{minipage}
\end{center}
\end{figure}

In a turn (see figure~\ref{fig5}), the centrifugal inertial force $\vC$,
\begin{equation}
\label{defFC} 
\vC = \pm \, m \, \frac{v^2}{R} \,
\frac  {\vlat} {\|\vlat\|}\,,
\end{equation}
has to be added to or subtracted from
the lateral gravitational force $\vlat = \vF_S - \vF_P$ to obtain the
total radial force, which we name $\vLAT$ in order to 
differentiate it from $\vlat$. In the second (``lower'') 
half of a left as well as in the second (``lower'')
half of a right curve,
the centrifugal force is parallel to the lateral force,
and the positive sign prevails in equation (\ref{defFC}).
By contrast, in the upper half of either a curve,
the centrifugal force is antiparallel to the lateral force,
and the negative sign should be chosen.
We have (see equations (\ref{defFN}), (\ref{defFS})~and~\ref{defFlat}))
\begin{equation}
\label{defFLAT}
\vLAT = \vlat + \vC\,.
\end{equation}	
The force $\vlat$ may be parallel or antiparallel 
to the centrifugal force, resulting in a different formula
for $\vLAT$ in each case,
\begin{equation}
\| \vLAT \| = \left| m \frac{v^2}{R} \pm
m\, g \sin\alpha \, |\cos\beta| \right| \,.
\end{equation}	
The new effective weight is
\begin{equation}
\label{defLOAD}
\vLOAD = \vN + \vLAT\,.
\end{equation}
The tilt angle changes as we replace $\vLoad \to \vLOAD$.
We denote the new tile angle by $\Phi$ as opposed to
$\phi$ (see equation~(\ref{defphi})).
Using a relation analogous to (\ref{cosphi}),
\begin{equation}
\label{cosphi2}
\cos\Phi = \frac{\vLOAD \cdot \vN }{\| \vLOAD \| \, \| \vN \|} \,,
\end{equation}
or by an elementary geometrical consideration (using the 
orthogonality of $\vLAT$ and $\vN$), we obtain
\begin{equation}
\label{tanPhi}
\tan\Phi = \left| \frac{v^2}{g \, R \, \cos \alpha} \,\pm \, 
\tan\alpha\, | \cos\beta | \right| \,.
\end{equation}
Furthermore, we assume the skier's velocity to be sufficiently
large that the centrifugal force dominates the lateral force; 
in this case we have
\begin{equation}
m \, \frac{v^2}{R} > m\, g \sin\alpha \, |\cos\beta| \qquad \mbox{and} \qquad
\frac{v^2}{g \, R \, \cos \alpha} > \tan\alpha\, | \cos\beta | 
\end{equation}
along the entire curve.
In this case, for a right curve, with $\beta = 0^o$ at the outset
and $\beta = 180^o$ at the end of the curve, and with 
$\cos\beta$ changing sign at $\beta = 90^o$, the correct
sign in equation (\ref{tanPhi}) is 
\begin{equation}
\label{fastright}
\tan\Phi = \frac{v^2}{g \, R \, \cos \alpha} -
\tan\alpha\, \cos\beta \qquad
\mbox{(fast skier, right curve)}\,.
\end{equation}

%
%

%
\section{The Ideal--Carving Equation}
\label{sICE}

%
%
\subsection{Derivation}
\label{ssDerivation}

Up to this point we have mainly followed the discussion
outlined on pp.~76--104 and pp.~208--215 of~\cite{LiSa1996}.
Under the assumption of a perfect carved turn,
the instantaneous curvature radius $R$, which is determined by 
the bent edges of the ski, depends on the sidecut radius $R_{\rm SC}$ 
and on the tilt angle $\Phi$ as follows (see equation (\ref{Rphi})),
\begin{equation} 
\label{RPhi}
R(\Phi) = R_{\rm SC} \, \cos\Phi\,.
\end{equation}
However, the assumption of a carved turn requires that the 
effective weight $\vLOAD$ be acting along the straight line
joining the ski boots and the center-of-mass of the skier.
This means that the angle $\Phi$ also has to fulfill
the equation (\ref{tanPhi}) which we specialize 
to the case (\ref{fastright}) in the sequel, 
\begin{equation} 
\label{ingredient1}
\tan\Phi = \frac{v^2}{g\,R\,\cos\alpha} \, - \, \tan\alpha\,\cos\beta\,.
\end{equation}
In view of (\ref{RPhi}), we have
\begin{equation} 
\label{ingredient2}
\tan\Phi = \sqrt{\frac{R_{\rm SC}^2}{R^2}-1}\,.
\end{equation}
Combining (\ref{ingredient1}) and (\ref{ingredient2}), we obtain
the {\em ideal-carving equation}
\begin{equation} 
\label{ICE}
\sqrt{\frac{R_{\rm SC}^2}{R^2}-1} = \frac{v^2}{g \, R \, \cos \alpha} - 
\tan\alpha\,\cos\beta\,.
\end{equation}
The variables
of the ideal-carving equation
are the velocity $v$ of the skier, the angle $\beta$ of the
trajectory of the skier with the horizontal, and the
instantaneous curvature radius $R$ of the skier's trajectory.
The parameters of (\ref{ICE})
are the inclination of the ski slope $\alpha$, the acceleration 
of gravity $g$, and the sidecut radius $R_{\rm SC}$ of the ski. 
Under appropriate substitutions,
this equation is equivalent to equation (T5.3) on p.~209 
of~\cite{LiSa1996}. In contrast to~\cite{LiSa1996}, the 
tilt angle is eliminated from the equation in our formulation.

Alternatively speaking, the ideal-carving equation
defines a function
\begin{subequations}
\label{surf}
\begin{equation}
\label{surfa}
f(v,R,\beta) = \frac{v^2}{g\,R\,\cos\alpha} - \tan\alpha\cos\beta - 
\sqrt{\frac{R_{\rm SC}^2}{R^2}-1} \,,
\end{equation}
so that the equation
\begin{equation}
\label{surfb}
f(v,R,\beta) = 0
\end{equation}
\end{subequations}
defines the ``ideal-carving surface'' as a 2-dimensional
imbedded in a three-dimensional space spanned 
by $v$, $R$, and $\beta$.

Likewise, we may consider the ideal-gas equation
\begin{subequations}
\begin{equation}
p \, V \, = \, N \, k \, T
\end{equation}
where $p$ is the pressure, 
$V$ denotes the volume, $N$ the number of atoms, 
$k$ the Boltzmann constant, and $T$ the absolute temperature.
The ideal-gas equation may trivially be rewritten as follows,
\begin{equation}
F(p,V,T) = \frac{pV}{NkT}\, - \, 1\, = \, 0\,.
\end{equation}
\end{subequations}
Of course, the equation $F(p,V,T) = 0$ then defines a
(two-dimensional) surface embedded in 
$\mathbbm{R}^3$. The ideal-gas equation entails
the idealization of perfect thermodynamic equilibrium, 
yet in realistic processes a gas volume will not 
always be in such a state. Nevertheless, in order
to avoid sub-optimal performance within a Carnot-like
process (or by analogy: in order to avoid 
frictional losses when skiing), one may strive to keep
the system as close to equilibrium as possible at all times.

%
%
\subsection{Graphical Representation}
\label{ssGraphical}

\begin{figure}[htb!]
\begin{center}
\begin{minipage}{13.0cm}
\begin{center}
\vspace*{1cm}
\includegraphics[width=0.7\linewidth]{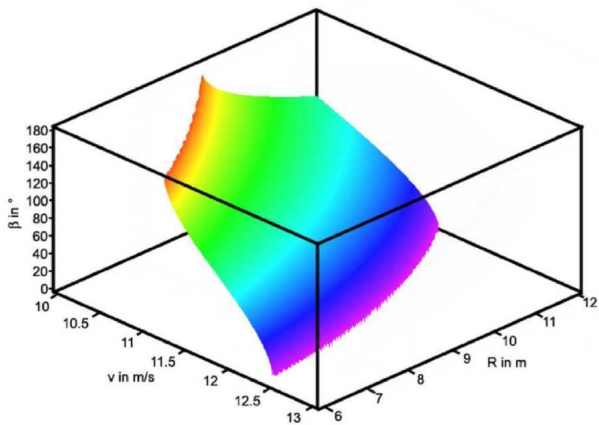}
\caption{\label{fig6}
The ideal-carving manifold for the parameters $R_{\rm SC} = 16~{\rm m}$, 
$g = 9.81~{\rm m}/{\rm s}^2$,
and $\alpha = 15^o$. The variables are the skier's
velocity $v$, the instantaneous radius of curvature $R$, and
the angle $\beta$ of the skier's trajectory with the 
horizontal.}
\end{center}
\end{minipage}
\end{center}
\end{figure}

The solutions of (\ref{surf}) define a 
(two-dimensional)
surface embedded in $\mathbbm{R} \times \mathbbm{R} \times [0^o, 180^o]$.
In figure~\ref{fig6},
the surface defined by equation (\ref{surfb}) is
represented for the parameter combination
$R_{\rm SC} = 16~{\rm m}$, $g = 9.81~{\rm m}/{\rm s}^2$,
and $\alpha = 15^o$.

%
%

Figure~\ref{fig6} gives us rather important information. 
In particular, we see that maintaining the ideal-carving condition
while going through a curve of constant radius of curvature,
within the interval $\beta = 0^o$ to $\beta = 180^o$,
implies a decrease in the skier's speed. This is possible only if
the frictional force, antiparallel to $\vP$,
provides for sufficient deceleration.
Of course, under ideal racing conditions the skier
will accelerate rather than decelerate during her/his descent.

%
%
\begin{figure}[htb!]
\begin{center}
\begin{minipage}{13.0cm}
\begin{center}
\vspace*{1cm}
\includegraphics[width=0.7\linewidth]{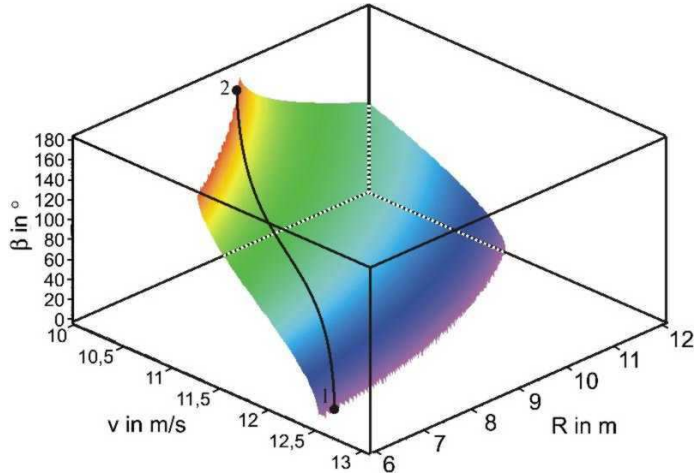}
\caption{\label{fig7} 
A hypothetical trajectory of a skier maintaining the ideal-carving
condition (\ref{surf}) during a right turn. Parameters are the 
same as in figure~\ref{fig6}. Note that the skier's velocity 
decreases from $v \approx 12.5\,{\rm m}/{\rm s}$ to
$v \approx 10.5\,{\rm m}/{\rm s}$ during the turn.}
\end{center}
\end{minipage}
\end{center}
\end{figure}

Acceleration during a turn is compatible with the
ideal carving condition only if the instantaneous
radius of curvature significantly decreases during
the turn. This corresponds to a turn in which
the skier starts off with a very wide turn, gradually
making the turn more tight during his/her descent.
The pattern generated is that of the letter ``J'',
and the corresponding curve is therefore commonly
referred to as a ``J-curve''~\cite{LiSa1996}.
The practical necessities of world-cup racing prevent 
such trajectories. Typical tilt angles are too high,
and typical centrifugal forces too large to be sustainable
in practice on the idealized trajectories.
This is why we see snow spraying even in highly 
competitive world-cup slalom and giant slalom skiing.

%
%
\begin{figure}[htb!]
\begin{center}
\begin{minipage}{12.0cm}
\begin{center}
\vspace*{1cm}
\includegraphics[width=0.7\linewidth]{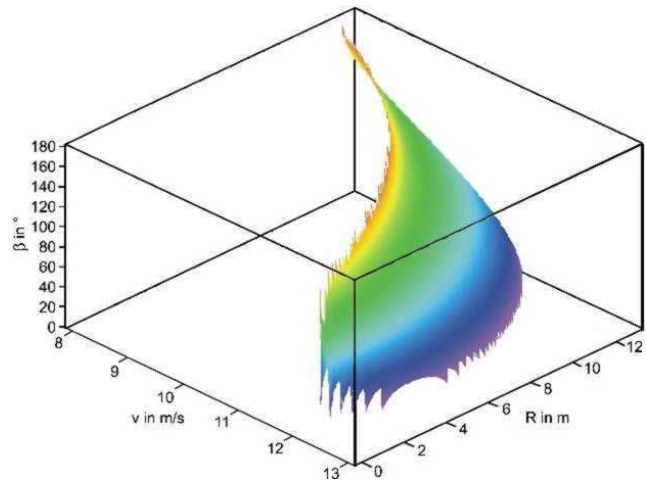}
\caption{\label{fig8} 
The ideal-carving manifold in the range of small $R$.
Parameters are the same as figures~\ref{fig6}
and~\ref{fig7}.}
\end{center}
\end{minipage}
\end{center}
\end{figure}

However, there is yet another very important restriction
to the possibility of maintaining ideal-carving conditions
at all times: An inspection of 
figures~\ref{fig6} and~\ref{fig7}
suggests that for an appreciable radius of curvature,
the velocity $v$ decreases monotonically
with the radius, $\beta$ being held constant. 
We now investigate the specific
velocity compatible with ideal-carving conditions 
at very tight turns $R \to 0$. The tilt angle $\Phi$ tends
to values close to $90^o$ in this case, because
\begin{equation}
\lim_{R\rightarrow 0} \tan\Phi = \lim_{R\rightarrow 0}
\sqrt{\frac{R_{\rm SC}^2}{R^2}-1} \to \infty \qquad
\mbox{as} \qquad R \to 0\,.
\end{equation}
Solving equation~(\ref{ICE}) for $v$, we obtain 
\begin{equation}
v(R,\beta) = \sqrt{g\,R\,\cos\alpha}\,
\sqrt{ \sqrt{\frac{R_{\rm SC}^2}{R^2}-1} + 
\tan\alpha\cos\beta } \to
\sqrt{g\,R_{\rm SC}\,\cos\alpha}
\qquad
\mbox{as} \qquad R \to 0\,.
\end{equation}
This latter relation holds independent of $\beta$, and 
this virtual independence of $\beta$ is represented graphically
in figure~\ref{fig8} for small $R$. As suggested by figures~\ref{fig6}
and~\ref{fig7}, it is impossible to maintain ideal-carving conditions
if the skier's velocity considerably exceeds the limiting velocity 
\begin{equation}
\label{vlimit}
v_{\rm limit} =  \sqrt{g\,R_{\rm SC}\,\cos\alpha}\,.
\end{equation}
We investigate this question in more detail.
For given $R$, the maximum $v$ is attained for 
$\beta=0$ because in this case, the centrifugal force
is most effectively compensated by the lateral force
$\vlat$. In this case,
\begin{eqnarray}
\label{taylor}
v(R,0) &=& \sqrt{g\,R\,\cos\alpha}\,
\sqrt{ \sqrt{\frac{R_{\rm SC}^2}{R^2}-1} + \tan\alpha } \nonumber\\[2ex]
&=& v_{\rm limit} \, \left(1 + \tan(\alpha)\,
\left(\frac{R}{R_{\rm SC}}\right)
- \frac{2 + \tan^2(\alpha)}{8}\,\left( \frac{R}{R_{\rm SC}} \right)^2 +
{\cal O}(R^2)\right) \,.
\end{eqnarray}
The maximum $v$ for which the ideal-carving condition 
can possibly be fulfilled is then determined by
the condition 
\begin{equation}
\left. \frac{\partial v}{\partial R}\right|_{\beta = 0} = 0\,,
\end{equation}
which, upon considering the first two nonvanishing terms 
in the Taylor expansion (\ref{taylor}) for small $R/R_{\rm SC}$, 
leads to
\begin{subequations}
\begin{equation}
\label{approx}
R_{\rm max} \approx 
\frac{2 \, \tan\alpha}{2 + \tan^2\alpha} \, R_{\rm SC} = 
R_{\rm SC}\, \left( \alpha - \frac{\alpha^3}{6} + 
{\cal O}(\alpha^5) \right)\,,
\end{equation}
a result which happens to be exact up to the order of $\alpha^5$.
The exact solution is surprisingly simple,
\begin{equation}
\label{exact}
R_{\rm max} = R_{\rm SC}\, \sin\alpha\,.
\end{equation}
\end{subequations}

The maximum velocity compatible with ideal-carving conditions
is independent of $\alpha$,
\begin{equation}
\label{vmax}
v_{\rm max} = v(R_{\rm max}, 0) = \sqrt{g\,R_{\rm SC}}\,.
\end{equation}
It implies a tilt angle $\Phi = 90^o - \alpha$ at
$\beta = 0$. The velocity $v_{\rm max}$, of course,
equals the velocity of a body on a circular trajectory
of radius $R_{\rm SC}$ with the centrifugal force 
being compensated by an acceleration of magnitude $g$ toward
the center. Observe, however, that in the current case the instantaneous
radius of curvature is $R_{\rm SC}\, \sin\alpha$.

A numerical example: For $R_{\rm SC} = 14\,{\rm m}$, and
$\alpha = 30^o$, we have $R_{\rm max} = 7\,{\rm m}$
and $v_{\rm max} = 11.72\,{\rm m}/{\rm s}$.
The limiting velocity deviates by about $10\,\%$ and is given by
$v_{\rm limit} = 10.91\,{\rm m}/{\rm s}$.

\begin{figure}[htb!]
\begin{center}
\begin{minipage}{12.0cm}
\begin{center}
\vspace*{1cm}
\includegraphics[scale=0.6]{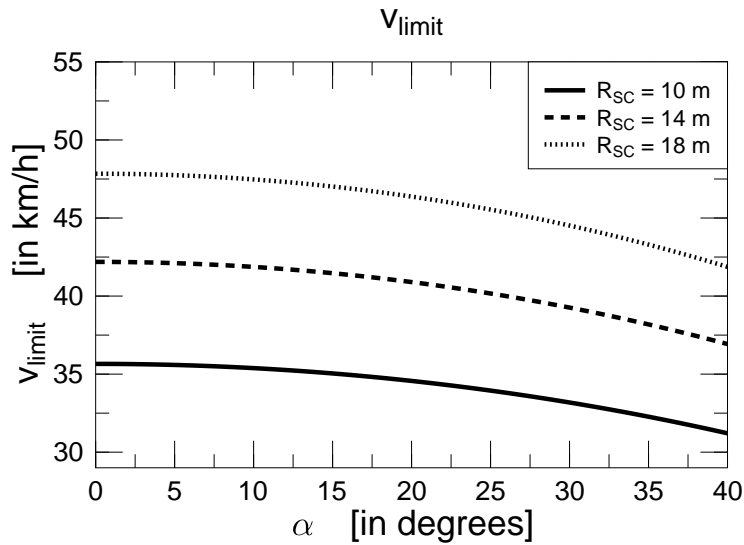}
\caption{\label{fig9}
The limiting velocity $v_{\rm limit}$
[Eq.~(\ref{vlimit})] is displayed 
for typical inclinations of the ski slope $\alpha$.
For velocities appreciably beyond $v_{\rm limit} \approx v_{\rm max}$
[see Eqs.~(\ref{vlimit}) and~(\ref{vmax})], it is 
impossible to maintain ideal-carving conditions.}
\end{center}
\end{minipage}
\end{center}
\end{figure}

Both the limiting velocity $v_{\rm limit}$
as well as the maximum velocity $v_{\rm max}$ 
are given purely as a function of 
the parameters of the ideal-carving equation (\ref{ICE}):
these are the acceleration of gravity $g$,
the sidecut radius $R_{\rm SC}$, and the 
inclination of the ski slope $\alpha$.
Furthermore, we observe that for typical parameters
as given in figure~\ref{fig8}, the ideal velocity 
of the skier varies only within about $30\,\%$ for all
radii of curvature in the interval $[0 \, {\rm m}, 12\,{\rm m}]$,
and all possible $\beta$.
That is to say, the limiting $v_{\rm limit}$ also gives
a good indication of the velocity range under which a carving 
ski, or a snowboard, can operate under nearly ideal-carving 
conditions (see also figure~\ref{fig9}).

%
%

%
%
\section{Implications and Conclusions}
\label{sConclusions}

In section~\ref{sToward}, we have discussed in detail
the forces acting on a skier during a carved turn,
as well as basic geometric properties of carving skis
and snowboards. In section~\ref{sICE},
We have discussed in detail the derivation of
the ideal-carving equation (\ref{ICE}) which 
establishes a relation between the skier's velocity,
the radius of curvature of the skier's trajectory and
the angle of the skier's course with the horizontal.
This equation determines an ideal-carving manifold whose properties
have been discussed in section~\ref{ssGraphical}, with 
graphical representations for typical parameters
to be found in figures~\ref{fig6}---\ref{fig9}.
In particular, the limiting velocity $v_{\rm limit}$ 
as given in equation (\ref{vlimit}) indicates 
an ideal operational velocity of a carving ski as a 
function of the angle of inclination of the ski slope
and of the sidecut radius.

The range of the limiting
velocities indicated in figure~\ref{fig9}
are well below those attained in world-cup downhill skiing.
In downhill skiing, the usage of skis with an appreciable 
sidecut is therefore not indicated. 
However, the velocity range of figure~\ref{fig9}
is well within the typical values attained in 
slalom races. It is therefore evident that 
carving skis are well suited for such races, 
in theory as well as in practice.
A slalom with tight turns, 
which implies a rather slow operational velocity due to 
the necessity of changing the trajectory within the reaction
time of a human being, demands slalom skis with a smaller
sidecut radius than those suited for a rather flat slope
and wide turns. Note that a smaller sidecut radius
implies a larger actual sidecut $d$
according to equation (\ref{RSC}). It may well be beneficial for a slalom 
skier to have a look at the actual course, and to measure 
the average steepness of the slope, and to choose an 
appropriate ski from a given selection, before starting
her/his race.

We will now discuss possible further improvements in the design 
of carving skis. To this end we draw an analogy to the steering of 
a bicycle traveling on an inclined surface. The driver is 
supposed not to exert any force via the action
of the pedals of the bicycle.
Indeed, during the ride on a bicycle, the driver
can maintain ideal-``carving'' conditions under rather 
general circumstances, avoiding slippage.
One might ask why a bicycle driver can 
accomplish this while a carving skier
or snowboarder cannot. The reason is the following:
Equation (\ref{RPhi}) defines a relation between the tilt 
angle $\Phi$ and the instantaneous radius of curvature 
$R$. When riding a bicycle, one may freely adjust 
the relation between $\Phi$ and $R$ via the steering. 
On carving skis, the 
position of the ``steering'' is always uniquely related 
to the tilt angle $\Phi$ by equation~(\ref{RPhi}). 
On a bicycle, it is possible to use a small steering angle 
even if one leans substantially
toward the center of the curve. On a carving ski, the ``steering angle''
automatically becomes large when the tilt angle 
is large, resulting in a small radius of curvature (again under 
the assumption of ``ideal-carving'' conditions). 
This circumstance eventually leads to the limiting velocity 
$v_{\rm limit}$
beyond which it is impossible to operate a carving ski 
under ideal-carving conditions, as represented by
equation~(\ref{vmax}). Beyond the limiting velocity, 
the non-fulfillment of the ideal-carving equation
is visible by spraying snow. By contrast, on a bicycle, it is 
possible to adjust the steering of the front 
wheel so that the radius of curvature
as defined by the relative inclination of the front and rear 
wheels, and the inclination $\Phi$ of the bicycle itself,
fulfill the ideal-carving equation~(\ref{ICE}). 

A carving ski that offers the possibility
of steering could be constructed 
with the help of an inertial measurement device (see e.g.~\cite{Ki2003}).
This device is supposed to continuously read  
the tilt angle $\Phi$, the velocity of the 
skier $v$ and the instantaneous radius of curvature $R$,
as well as the angle $\beta$ 
and the inclination of the ski slope $\alpha$.
According to equation (\ref{ICE}), these variables 
determine an ideal sidecut radius $R_{\rm SC}$ 
which could be adjusted dynamically by a servo motor.
In this case, it would be possible to fulfill near-ideal-carving 
conditions along the entire trajectory.
A first step in this direction would be simpler device 
that measures only the inclination of the ski slope 
$\alpha$ and determines a near-ideal sidecut radius 
according to equation~(\ref{vlimit}).

%
%
\section*{Acknowledgements}

The authors thank Sabine Jentschura and Hans-Ulrich Fahrbach for carefully 
reading the manuscript, and for helpful discussions.

\end{document}